\pdfoutput=1
\documentclass[iop]{emulateapj}
\usepackage{apjfonts}
\usepackage{graphicx}

\newcommand{\be}{\begin{equation}}
\newcommand{\ee}{\end{equation}}
\newcommand{\bea}{\begin{eqnarray}}
\newcommand{\eea}{\end{eqnarray}}
\newcommand{\p}{\partial}

\let\phi=\varphi
\let\rho=\varrho

\shorttitle{Charged tori in spherical gravitational and dipolar magnetic fields}
\shortauthors{Slan\'{y} et al.}
\submitted{Version accepted for publication in ApJS.}

\begin{document}

\title{Charged tori in spherical gravitational and dipolar magnetic fields}

\author{P. Slan\'{y}, J. Kov\'{a}\v{r}, Z. Stuchl\'{\i}k}
\affil{Institute of Physics, Faculty of Philosophy and Science, Silesian University in Opava\\
Bezru\v{c}ovo n\'{a}m. 13, CZ-746\,01 Opava, Czech Republic, petr.slany@fpf.slu.cz}

\and

\author{V. Karas}
\affil{Astronomical Institute, Academy of Sciences,\\ 
Bo\v{c}n\'{\i} II, Prague, CZ-141\,31, Czech~Republic}

\begin{abstract}
A Newtonian model of non-conductive, charged, perfect fluid tori orbiting in combined spherical gravitational and dipolar magnetic fields is presented and stationary, axisymmetric toroidal structures are analyzed. Matter in such tori exhibits a purely circulatory motion and the resulting convection carries charges into permanent rotation around the symmetry axis. As a main result, we demonstrate the possible existence of off-equatorial charged tori and equatorial tori with cusps enabling outflows of matter from the torus also in the Newtonian regime. These phenomena represent qualitatively a~new consequence of the interplay between gravity and electromagnetism. From an astrophysical point of view, our investigation can provide insight into processes that determine the vertical structure of dusty tori surrounding accretion disks.
\end{abstract}



\section{Introduction}
As demonstrated by a~variety of observational phenomena ranging from spectral and timing properties of accreted gas in different energy bands, luminosity profiles, and velocity dispersions of stars forming the nuclear clusters, up to direct measurements of stellar motion in the Galactic center \citep{Krolik:2004:NATURE:,Eck-Scho-Str:2005:BlackHoleCenterMilkyWay:}, supermassive black holes are frequently present in the nuclei of galaxies. The radio morphology of jets emerging from galactic nuclei and the skewed double-horn profiles of many X-ray and UV spectral lines suggest that these objects obey axial rather than spherical symmetry \citep{Urr-Pad:1995:PASP:}. As a working hypothesis, a central black hole with a~radiating inner accretion disk is thought to be surrounded further out by a torus of obscuring material (dust).

Typical masses of supermassive black holes fall within the range of $M_\bullet\simeq10^6$--$10^8\,M_\odot$. The central mass is related to the gravitational radius of the black hole, $R_{\rm{}g}=GM_{\bullet}/c^2\simeq
1.5\times10^{13}(M_\bullet/10^8M_\odot)\;{\rm{}cm}\simeq
10^{-5}\;{\rm{}pc}$. On the other hand, the typical radius of the torus, as derived from observations, greatly exceeds the gravitational radius, $R_{\rm{}d}\simeq10$--$100$~pc.
The mass of the torus is orders of magnitude smaller than the black hole mass, $M_{\rm{}d}\ll M_\bullet$. Therefore, $M_{\rm{}d}$ is usually neglected in discussions, however, it may be important for the internal structure of the torus. In fact, the dusty tori are thought to be on the verge of self-gravitational instability, which can lead to the fragmentation of the torus when Toomre's criterion of stability is violated \citep{Col-Zah:2008:ASTRA:}.

Different types of active galactic nuclei in Seyfert galaxies can be unified by introducing some form of obscuring tori, which are believed to encircle the central black hole \citep{Ant-Mil:1985:ASTRJ2:,Urr-Pad:1995:PASP:}. The presence of a geometrically and optically thick dusty structure is an essential component of the unification scheme \citep{Hon-Kis:2010:ASTRA:}. The torus structure is thought to be inhomogeneous, in the form of molecular/dusty clumps (clouds). The matter forming the torus follows a common bulk orbital motion and its rotation defines the symmetry axis of the system, which may or may not be aligned with the rotation axis of the black hole. It is generally expected that the alignment is enforced at least in the inner regions by the Bardeen-Petterson effect \citep{Kum-Pri:1985:MONNR:}.

Equilibrium figures of gaseous tori have been studied in great detail \citep{Koz-Jar-Abr:1978:ASTRA:,Abr-Jar-Sik:1978:ASTRA:,Kat-Fuk-Min:2008:BHAccDis:}, however, the vertical component of the pressure gradient, required to maintain the equilibrium, does not seem to be sufficient in dusty tori (see, e.g., \citet{Mur-Yaq:2009:MONNR:} and references cited therein). Therefore, it appears that these structures have a tendency to collapse toward the plane. Moreover, the self-gravity influences the structure of massive tori (as well as geometrically thin Keplerian disks) once the local mass density of the accreted gas at a given location exceeds $\sim\Omega^2/4\pi G$, see, e.g., \citet{Shl-Beg:1987:NATURE:,Hur:1998:ASTRA:}. Despite the fact that signatures of obscuration (especially those seen in X-ray spectra) and variability properties strongly indicate the need for a significant vertical extent of obscuring tori in many Seyfert type 2 galaxies, the physical model for the tori remains uncertain. 

\begin{figure}
\centering
  \includegraphics[width=1 \hsize]{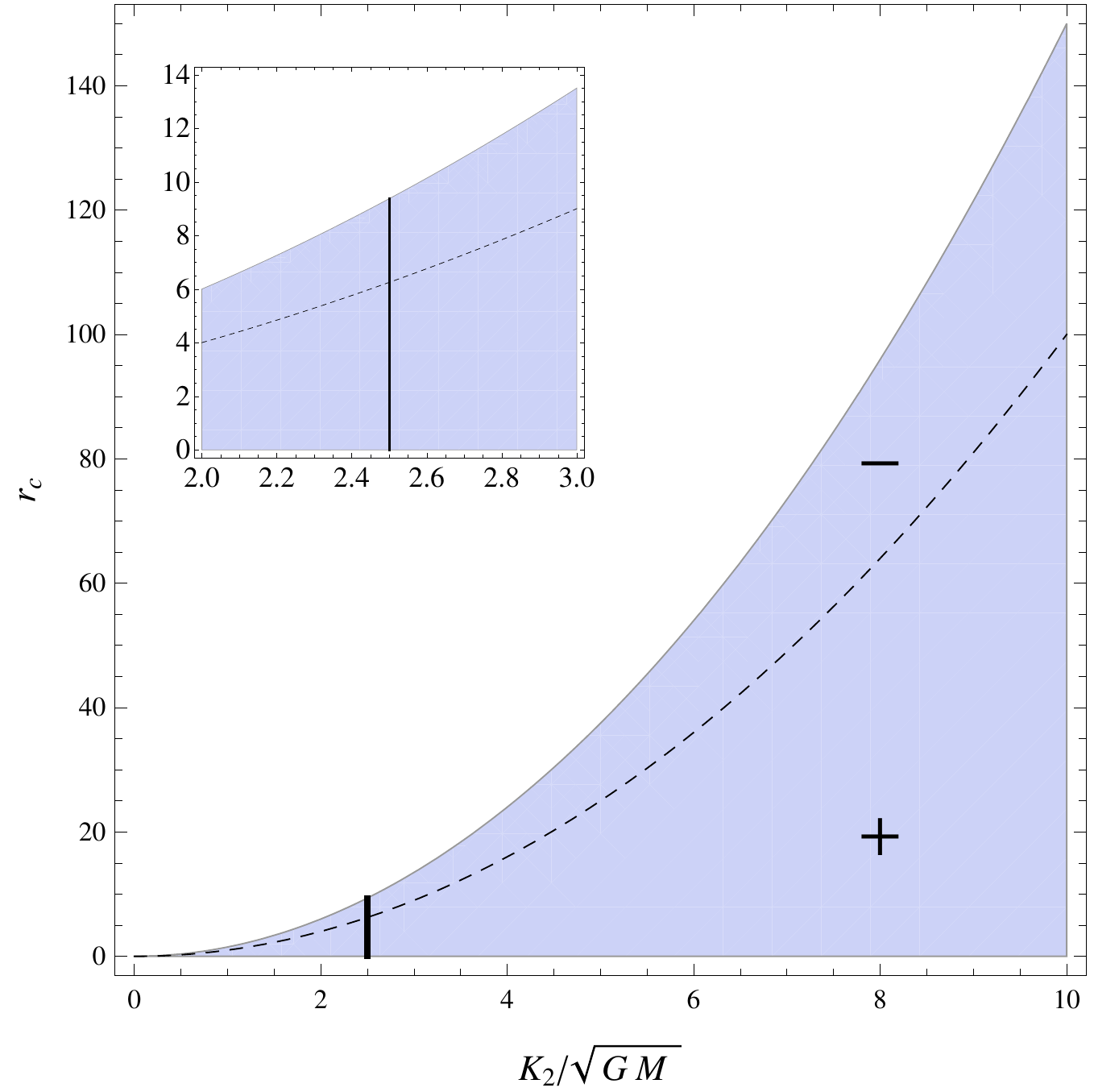}
  \caption{Region of parameters $K_2$ and $r_{\rm c}$, enabling the existence of family I and II equatorial tori with a~uniform distribution of the specific angular momentum, $\ell(r,\,\theta)=K_2=\mbox{const}$. The dashed curve divides the region into subregions with positively/negatively ($+/-$) charged tori. The interval of possible values of $r_{\rm c}$ for the choice $K_{2}/\sqrt{GM}=2.5$ is marked and enlarged.}
  \label{f1}
\end{figure}

The above-given arguments suggest that the vertical structure of dusty tori needs further discussion. For example, it has been proposed recently \citep{Cze-Hry:2011:ASTRA:} that vertical motions of the dust clumps play an important role. These can be driven by radiation 
pressure in the region where dust can survive evaporation. In fact, the irradiation in active galactic nuclei is very intense and it presumably leads to the dust evaporation below a critical radius (typically at a fraction of parsec) where the dust temperature reaches about $10^3$K. 

It is important to note that the irradiation of dust particles should produce a positive net electrical charge by photoionization \citep{Hor:1996:ARASTRA:,Vla-etal:2005:DustConvPlasma:}. On the other hand, plasma electron and ion currents are continuously entering the grain surface, so the sign and magnitude of the equilibrium charge depend on the total currents that are absorbed and emitted from the grain surface. This is a complex process that depends on many parameters; simultaneously, it appears to be an important mechanism that can govern the torus structure. The electrostatic charge is one of the essential parameters that controls the dynamics of dust grains embedded in the surrounding cosmic plasma \citep{Ish:2007:JPHYSD:}. 

When the electromagnetic force, due to the pervasive external magnetic field, is taken into account, an electrically charged dust can establish vertically extended structures that  `levitate' above and under the equatorial plane. Namely, charged test particles can be bounded within off-equatorial wells of effective potential, which suggests the existence of very dilute halo toroidal formations consisting of non-interacting, charged grains, ions, or electrons. The Newtonian study of charged dust grains orbiting in planetary magnetospheres and forming halo orbits was published, e.g., in \citet{How-Hor-Stu:1999:PHYRL:,Dul-Hor-How:2002:PHYSD:}, while the question of whether such halo orbits can appear also in strong gravitational fields near compact objects was positively answered in \citet{Kov-Stu-Kar:2008:CLAQG:,Stu-Kov-Kar:2009:IAUS:,Kov-etal:2010:CLAQG:}. Of course, in many astrophysical scenarios such simple test-particle approaches fail because of higher densities of charged matter in reality. Then, possible approaches follow from the kinetic theory (suitable for lower density matter) or from the hydrodynamics (suitable for higher densities). 

In this paper, we apply the Newtonian hydrodynamical approach, modeling perfect fluid tori with electric charge spread through the fluid of infinite resistivity. This model of dielectric fluid represents an opposite limit to the well-known ideal magnetohydrodynamics with zero resistivity as a good approximation of many astrophysical plasmas, see, e.g., \citet{Pun:2001:BlackHoleGravitohydromagnetics:}. The general relativistic description of charged perfect-fluid disks of infinite resistivity was published recently in \citet{Kov-etal:2011:PHYSR4:}, where the tori encircling static electrically charged Reissner-Nordstr{\o}m black hole were analyzed. Note, however, that the introduced Newtonian model is basically (technically) simpler and more convenient for our primary investigations of halo tori formed far away from the central black hole (i.e., in a~weak gravitational field) due to the presence of external magnetic fields.

We approach the problem within the framework of a two-dimensional (axially symmetric) toy model of electrically charged tori rotating in spherical gravitational and dipolar magnetic 
fields. By constructing the equilibrium figures, we find the constraints on the spatial distribution of charge and matter densities, and on the corresponding angular momentum profile. The main aim of our investigation is to explore how the well-known structures of electrically neutral gaseous tori become modified by the interaction of (small) charge density with the poloidal magnetic field, leading to vertically extended steady distribution which can reach further out of the equatorial plane and remain in a~stable permanent azimuthal circulation.

The paper is organized as follows. The equation of motion for a perfect fluid with spatially distributed electric charge (Euler equation) and the corresponding condition of hydrostatic equilibrium are formulated in Sec.~\ref{s2}. In Sec.~\ref{s3}, this condition is analyzed for incompressible fluid and possible stationary toroidal configurations are found. The relevance of the results obtained for a polytropic fluid is discussed in Sec.~\ref{s4}. In Sec.~\ref{s5}, the use of Newtonian formalism for compact objects is demonstrated by replacing the Newtonian gravitational potential with the Paczy\'{n}ski--Wiita one and the corresponding modifications of isobars are discussed. Concluding remarks are given in Sec.~\ref{s6}.

\section{Newtonian model} \label{s2}
The equation of motion for a perfect fluid (no viscosity) orbiting in gravitational and electromagnetic fields, i.e., Euler's equation \citep{Lan-Lif:1987:FluidMechanics:}, has the form
\be                                                                  \label{e1}
\rho_{\rm m}(\p_t v_{i}+v^{j}\nabla_j v_{i})=-\nabla_{i}P-\rho_{\rm m}\nabla_{i}\Phi + \rho_{e}(E_{i}+\epsilon_{ijk}v^{j}B^{k}), 
\ee
where $\rho_{\rm m}$ and $\rho_{\rm e}$ are the mass-density and charge-density, respectively, $P$ denotes the pressure, $\mathbf{v}$ is the velocity field in the fluid, and $\Phi$ corresponds to the gravitational potential.
An electromagnetic field is described by its electric part $\mathbf{E}$ and magnetic part $\mathbf{B}$. The last term on the rhs of Euler's equation (\ref{e1}) thus corresponds to the Lorentz force density. 

\begin{figure*}
\centering
  \includegraphics[width=1\hsize]{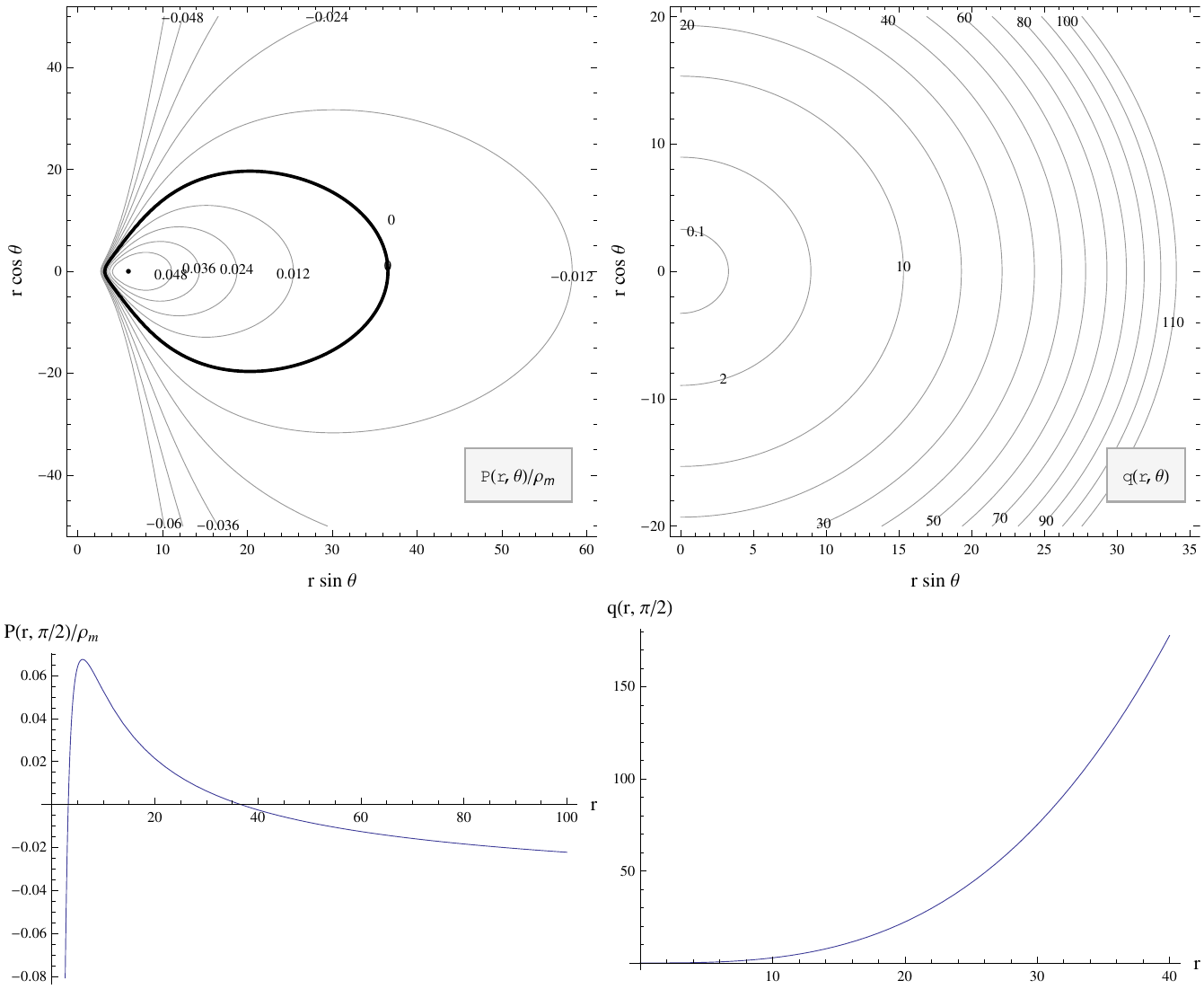}
	\caption{Positively charged equatorial torus (\emph{family I}). The meridional section and equatorial profile of the pressure field (left) and the specific electric charge (right) are shown.}
  \label{f2}
\end{figure*}

In our analysis, we assume stationary and axisymmetric flow of test charged perfect fluid in external spherical gravitational and dipolar magnetic fields. In spherical polar coordinates ($r,\,\theta,\,\phi$),
\bea                                                                 \label{e2}
\Phi &=& -\frac{GM}{r}, \\
                                                                     \label{e4}
E_{i} &=& 0, \quad i=(r,\,\theta,\,\phi), \\                             \label{e5}
B_{r} &=& 2\mu\frac{\cos\theta}{r^3}, \quad B_{\theta}=\mu\frac{\sin\theta}{r^3},
\eea
where $M$ is the mass of central object, $\ell(r,\,\theta)$ is the specific angular momentum of fluid elements, and $\mu>0$ corresponds to the magnetic dipole moment of the external magnetic field.

Conservations of mass and electric charge are described by corresponding continuity equations
\bea                                                                 \label{e6}
\p_t \rho_{\rm m}+\nabla_i (\rho_{\rm m}v^{i}) &=& 0, \\
\p_t \rho_{\rm e}+\nabla_i (\rho_{\rm e}v^{i}) &=& 0.                    \label{e7}
\eea
For stationary and axisymmetric flow,
\be                                                                  \label{e8}
v_{r}=v_{\theta} = 0, \quad
v_{\phi} = v_{\phi}(r,\,\theta),
\ee
thus, both the continuity equations (\ref{e6}) and (\ref{e7}) are fulfilled automatically. 

The proposed fluid dynamics is described by two partial differential equations, following from Euler's equation (\ref{e1}):
\bea                                                                 \label{e9}
\frac{\p P}{\p r} &=& -\rho_{\rm m}\frac{GM}{r^2}+\rho_{\rm m}\frac{v_{\phi}^2}{r}-\rho_{\rm e}v_{\phi}\mu\frac{\sin\theta}{r^3}, \\                              \label{e10}
\frac{1}{r}\frac{\p P}{\p\theta} &=& \rho_{\rm m}\frac{v_{\phi}^2}{r}\cot\theta + 2\rho_{\rm e}v_{\phi}\mu\frac{\cos\theta}{r^3}.
\eea
In order to solve this set of equations, it is useful to assume charge density in the form
\be                                                                  \label{e11}
\rho_{\rm e}=\rho_{\rm m}q(r,\,\theta),
\ee
where the function $q(r,\,\theta)$ describes the specific charge distribution in the fluid. To complete the set of equations, we need to establish an equation of state.

\begin{figure*}[ht]
\centering
  \includegraphics[width=1\hsize]{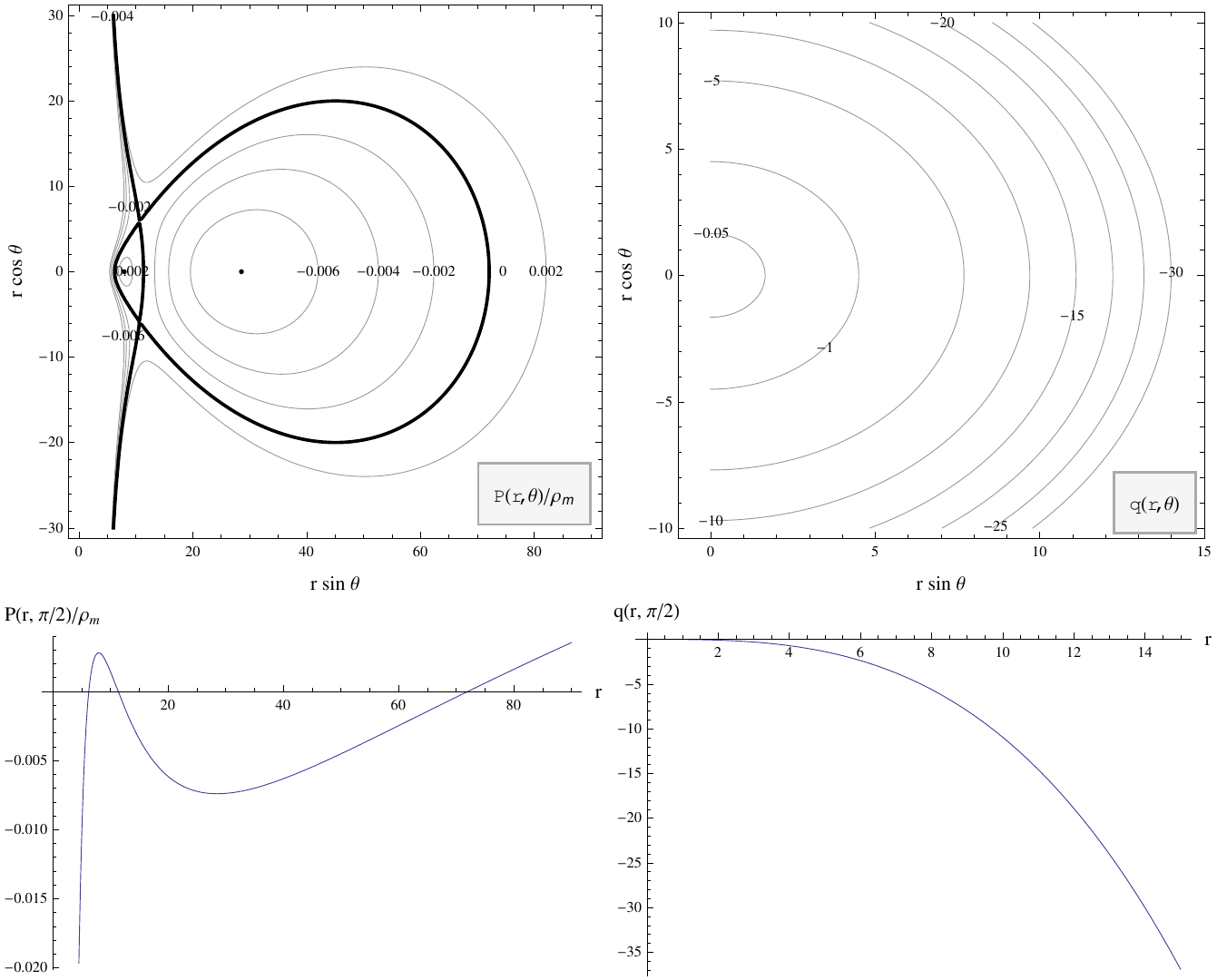}
	\caption{Negatively charged equatorial torus (\emph{family I}). The meridional section and equatorial profile of the pressure field (left) and the specific electric charge (right) are shown.}
  \label{f3}
\end{figure*}

\section{Incompressible fluid} \label{s3}
When there is no significant compression or expansion of the fluid (liquid or gas) anywhere along flow lines, we can use the limit of incompressible fluid, for which $\rho_{\rm m}=\mbox{const}$. Equations (\ref{e9}) and (\ref{e10}) then have the form
\bea                                                                 
\frac{1}{\rho_{\rm m}}\frac{\p P}{\p r} &=& \left(-\frac{GM}{r^2}+\frac{v_{\phi}^2}{r}-\mu\, q\,v_{\phi}\frac{\sin\theta}{r^3}\right)\equiv\mathcal{A}(r,\,\theta),  \label{e12} \\   
\frac{1}{\rho_{\rm m}}\frac{\p P}{\p\theta} &=& \left(v_{\phi}^2\cot\theta + 2\mu\, q\,v_{\phi}\frac{\cos\theta}{r^2}\right)\equiv\mathcal{B}(r,\,\theta).  \label{e13}
\eea
A necessary and sufficient condition for the existence of the solution of the above-mentioned set of partial differential equations, describing a distribution of pressure in the fluid, is given by the integrability condition
\be                                                                  \label{e14}
\frac{\p\mathcal{A}}{\p\theta}=\frac{\p\mathcal{B}}{\p r},
\ee
having the form
\bea                                                                 
\lefteqn{
2v_{\phi}\left(\frac{1}{r}\frac{\p v_{\phi}}{\p\theta}-\cot\theta\frac{\p v_{\phi}}{\p r}\right) + \mu\left[3q\,v_{\phi}\frac{\cos\theta}{r^3} -\frac{\sin\theta}{r^3}\right.}   \nonumber \\
& &  \left. \times\left(v_{\phi}\frac{\p q}{\p\theta}+q\frac{\p v_{\phi}}{\p\theta}\right) - 2\frac{\cos\theta}{r^2}\left(v_{\phi}\frac{\p q}{\p r}+q\frac{\p v_{\phi}}{\p r}\right)\right] =0.                                                 \label{e15}
\eea

\begin{figure*}[ht]
\centering
  \includegraphics[width=1 \hsize]{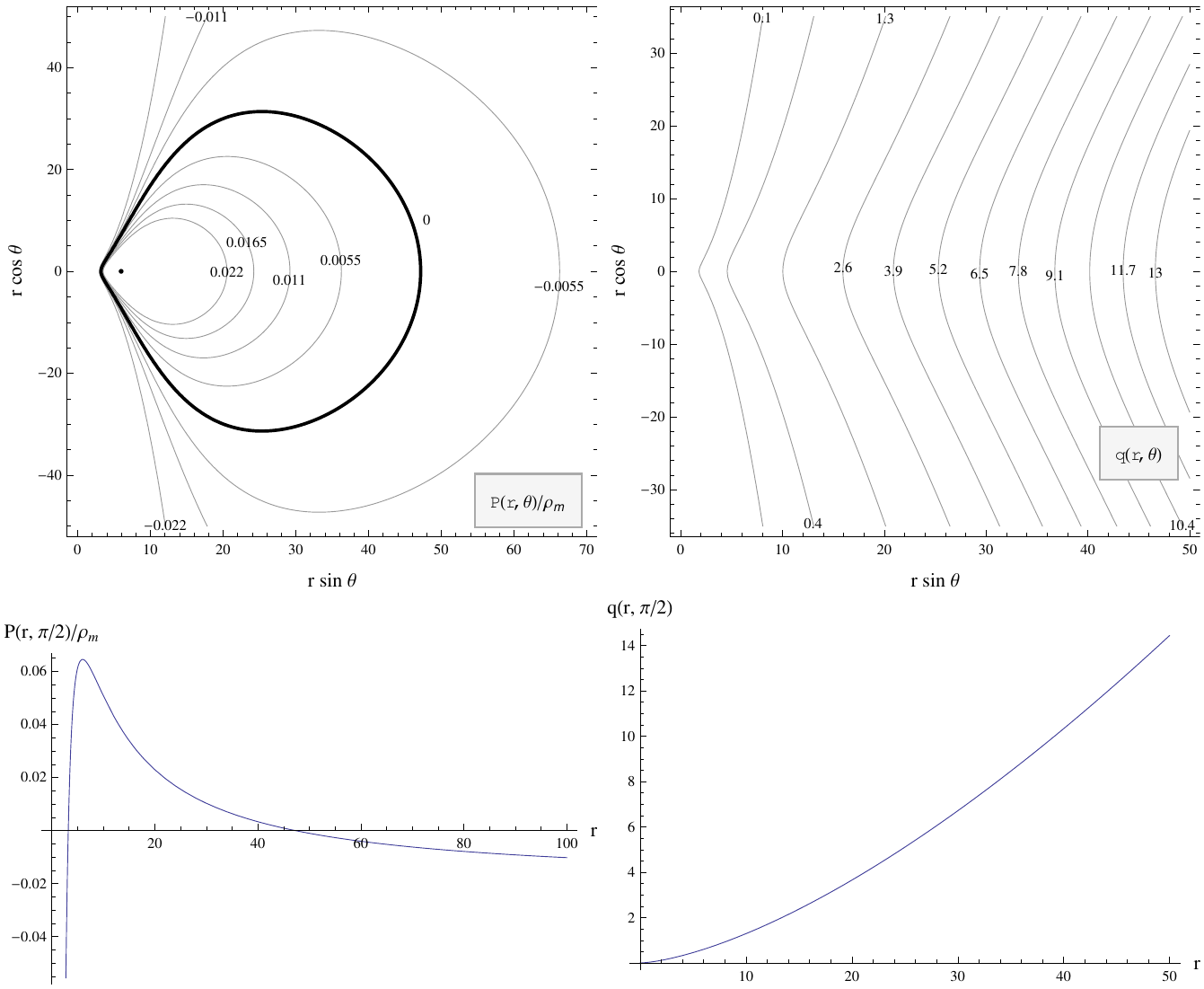}
  \caption{Positively charged equatorial torus (\emph{family II}). The meridional section and equatorial profile of the pressure field (left) and the specific electric charge (right) are shown.}
	\label{f4}
\end{figure*}

The integrability condition has the form of the partial differential equation for two unknown functions $v_{\phi}(r,\,\theta)$ and $q(r,\,\theta)$. In order to find their proper forms, it is necessary to specify somehow at least one of these functions. In this paper, we choose the function $v_{\phi}(r,\,\theta)$, i.e., the way of rotation of the fluid.
Analysis of condition (\ref{e15}) reveals that in the case of $q=0$ (or $\mu=0$) the orbital velocity $v_{\phi}$ can be written in a separate form $v_{\phi}(r,\,\theta)=v(r)u(\theta)$, where the functions $v(r)$ and $u(\theta)$ satisfy differential equations (ensuring that the first parenthesis in the integrability condition (\ref{e15}) is zero)
\be                                                                  \label{e16}
\frac{\mathrm{d}\ln v}{\mathrm{d}\ln r}=\frac{\mathrm{d}\ln u}{\mathrm{d}\ln(\sin\theta)}=K_1,
\ee
with $K_1$ being a separation constant. Solving Eqs. (\ref{e16}), we obtain a general form of the orbital velocity profile 
\be                                                                  \label{e17}
v_{\phi}(r,\,\theta)=K_2 (r\sin\theta)^{K_1},
\ee
where $K_2$ is an integration constant. 
We will use such general form of $v_{\phi}(r,\,\theta)$ also in situations, in which $q\neq 0$ (and $\mu\neq 0$).

Further, we assume that also the distribution of a specific charge in the fluid, i.e., the function $q(r,\,\theta)$, can be given in the separable form $q(r,\,\theta)=q_1(r)\,q_2(\theta)$. Then there are four general forms of such distributions satisfying the integrability condition (\ref{e15}):
\bea
q(r,\,\theta) &=& Cr^{-3(K_{1}-1)/2},                                \label{e18} \\
q(r,\,\theta) &=& Cr^{3/2}(\sin\theta)^{-3K_{1}},                    \label{e19} \\
q(r,\,\theta) &=& Cr^{-3K_{1}/2}\sin^{3}\theta,                      \label{e20} \\
q(r,\,\theta) &=& C(\sin\theta)^{3(1-K_{1})};                        \label{e20.1}
\eea
$C$ is a constant of integration. Further, we shall refer to distributions (\ref{e18})--(\ref{e20.1}) as families I, II, III and IV, respectively. We can see that for certain values of constant $K_1$, there are some overlaps between those families: for $K_1=1$, families I and IV coincide, for $K_1=0$, families I, II as well as III, IV coincide, and for $K_1=-1$ this is true for families II and III.

In the center of the torus $(r_{\rm c},\theta_{\rm c})$, the pressure takes its maximal value. The necessary condition $\nabla P=0$ (especially $\p P/\p\theta=0$, see relation (\ref{e13})) points out that the pressure maxima can be located either in the equatorial plane $(\theta_{\rm c}=\pi/2)$ or even out of the equatorial plane $(\theta_{\rm c}\neq\pi/2)$, where the possible off-equatorial maxima are supported by the specific charge of the torus with its distribution along the off-equatorial circle given by the relation
\be                                                                  \label{e21}
q(r_{\rm c},\,\theta_{\rm c})=-\frac{v_{\phi}(r_{\rm c},\,\theta_{\rm c})}{2\mu}\frac{r_{\rm c}^2}{\sin\theta_{\rm c}}.
\ee

The condition $\nabla P=0$ is only a necessary condition for an existence of local extrema. To verify their real existence, we must evaluate the second order derivatives and construct the Hessian matrix
\be                                                                  \label{e22}
\mathcal{H} =
\left( 
\begin{array}[c]{cc}
\frac{\p^2 P}{\p r^2} & \frac{\p^2 P}{\p r\p\theta} \\
\frac{\p^2 P}{\p\theta\p r} & \frac{\p^2 P}{\p\theta^2}
\end{array}
\right);
\ee
in the loci of pressure maxima, two conditions must be satisfied:
\be                                                                  \label{e23}
\frac{\p^2 P}{\p r^2}|_{r=r_{\rm c},\theta=\theta_{\rm c}}<0,\quad
\det\mathcal{H}|_{r=r_{\rm c},\theta=\theta_{\rm c}}>0.
\ee

\begin{figure}
\centering
  \includegraphics[width=1 \hsize]{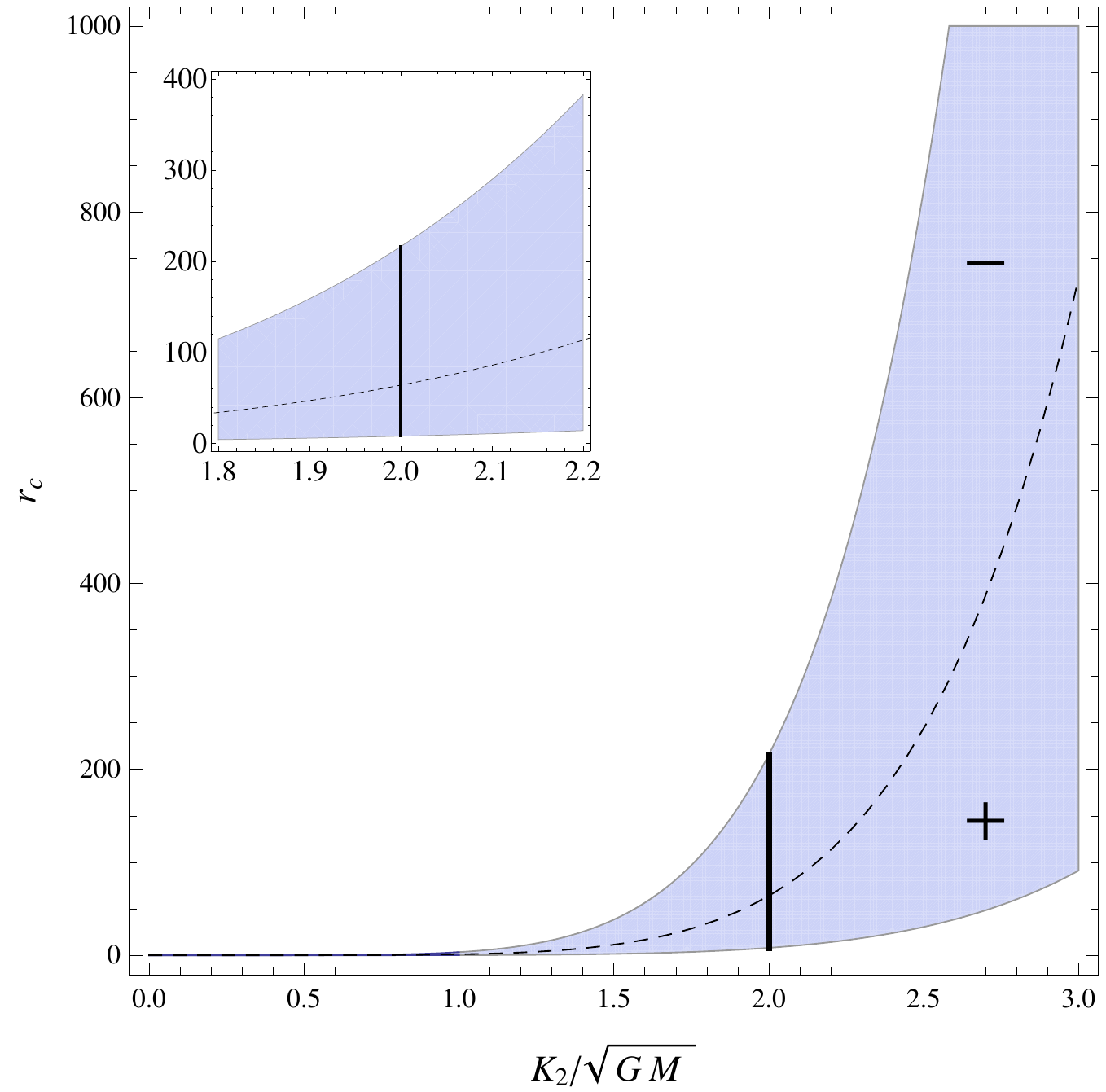}
  \caption{Region of parameters $K_2$ and $r_{\rm c}$, enabling the existence of family III equatorial tori with the specific angular momentum profile $\ell(r,\,\theta)=K_2(r\sin\theta)^{1/3}$. The dashed curve divides the region into subregions with positively/negatively ($+/-$) charged tori. The interval of possible values of $r_{\rm c}$ for the choice $K_{2}/\sqrt{GM}=2.0$ is marked and enlarged.}
  \label{f5}
\end{figure}

\subsection{Equatorial tori}
A local maximum of the pressure corresponding to the center of the torus is located in the equatorial plane, $\theta_{\rm c}=\pi/2$. From Eq.~(\ref{e12}), it follows that the orbital velocity $v_{\phi}(r,\,\theta)>0$ is given by the relation
\be                                                                  \label{e24}
v_{\phi}(r_{\rm c}) = \sqrt{\frac{GM}{r_{\rm c}}\left(1+\frac{\mu^{2}q^2}{4GMr_{\rm c}^3}\right)}+ \frac{\mu q}{2r_{\rm c}^2}
\ee
there.

Analysis of conditions (\ref{e23}) shows that equatorial local maxima of the pressure exist for all four families (\ref{e18})--(\ref{e20.1}) of the specific charge $q(r,\,\theta)$ if, in the center of torus, the constants $K_1$, $K_2$, and $C$ satisfy the conditions
\begin{description}
\item[family I]
\bea
\frac{K_{2}^2}{GM}r_{\rm c}^{2K_1+1} &\lessgtr& \frac{K_1-1}{5K_1+1}\quad 
\mbox{for}\ K_1\gtrless-\frac{1}{5},                                          \label{e25} \\
\frac{K_{2}^2}{GM}r_{\rm c}^{2K_1+1} &>& \frac{2}{3},               \label{e26} \\
\mu C &=& K_{2}r_{\rm c}^{(5K_1+1)/2}-\frac{GM}{K_2}r_{\rm c}^{(K_1-1)/2}, \label{e27}
\eea
\item[family II]
\bea
K_1 &<& -\frac{1}{2},                                                \label{e28} \\
\frac{K_{2}^2}{GM}r_{\rm c}^{2K_1+1} &>& \frac{2}{3},               \label{e29} \\
\mu C &=& K_{2}r_{\rm c}^{(2K_1+1)/2}-\frac{GM}{K_2}r_{\rm c}^{-(2K_1+1)/2}, \label{e30}
\eea
\item[family III]
\bea
\frac{K_{2}^2}{GM}r_{\rm c}^{2K_1+1} &\lessgtr& \frac{K_1+2}{5K_1+4} \quad
\mbox{for}\ K_1\gtrless-\frac{4}{5},                                \label{e31} \\
\frac{K_{2}^2}{GM}r_{\rm c}^{2K_1+1} &>& \frac{2}{3},               \label{e32} \\
\mu C &=& K_{2}r_{\rm c}^{(5K_1+4)/2}-\frac{GM}{K_2}r_{\rm c}^{(K_1+2)/2}, \label{e33}
\eea 
\item[family IV]
\bea
\frac{K_{2}^2}{GM}r_{\rm c}^{2K_1+1} &\lessgtr& \frac{1-K_1}{K_1+2} \quad
\mbox{for}\ K_1\gtrless -2,                                \label{e33.1} \\
\frac{K_{2}^2}{GM}r_{\rm c}^{2K_1+1} &>& \frac{2}{3},               \label{e33.2} \\
\mu C &=& K_{2}r_{\rm c}^{(K_1+2)}-\frac{GM}{K_2}r_{\rm c}^{(1-K_1)}. \label{e33.3}
\eea 
\end{description}

In the following paragraphs, we present illustrative examples of equatorial tori for each of the specific charge distribution families. A~particular solution of the set of equations (\ref{e12}) and (\ref{e13}) is obtained when the constants $K_1$, $K_2$, and $C$ are specified with respect to conditions (\ref{e25})--(\ref{e33.3}) for a~given family. Instead of the value of $C$, the location $r_{\rm c}$ of the torus center can be prescribed. Of course, the choice of constants should be physically meaningful, reflecting realistic ideas about the motion of matter. Such a basic idea is, e.g., subluminal motion inside the torus, $v_{\phi}<c$ ($c$ is the vacuum speed of light), which is not automatically incorporated in the Newtonian description.

\begin{figure*}
\centering
  \includegraphics[width=1\hsize]{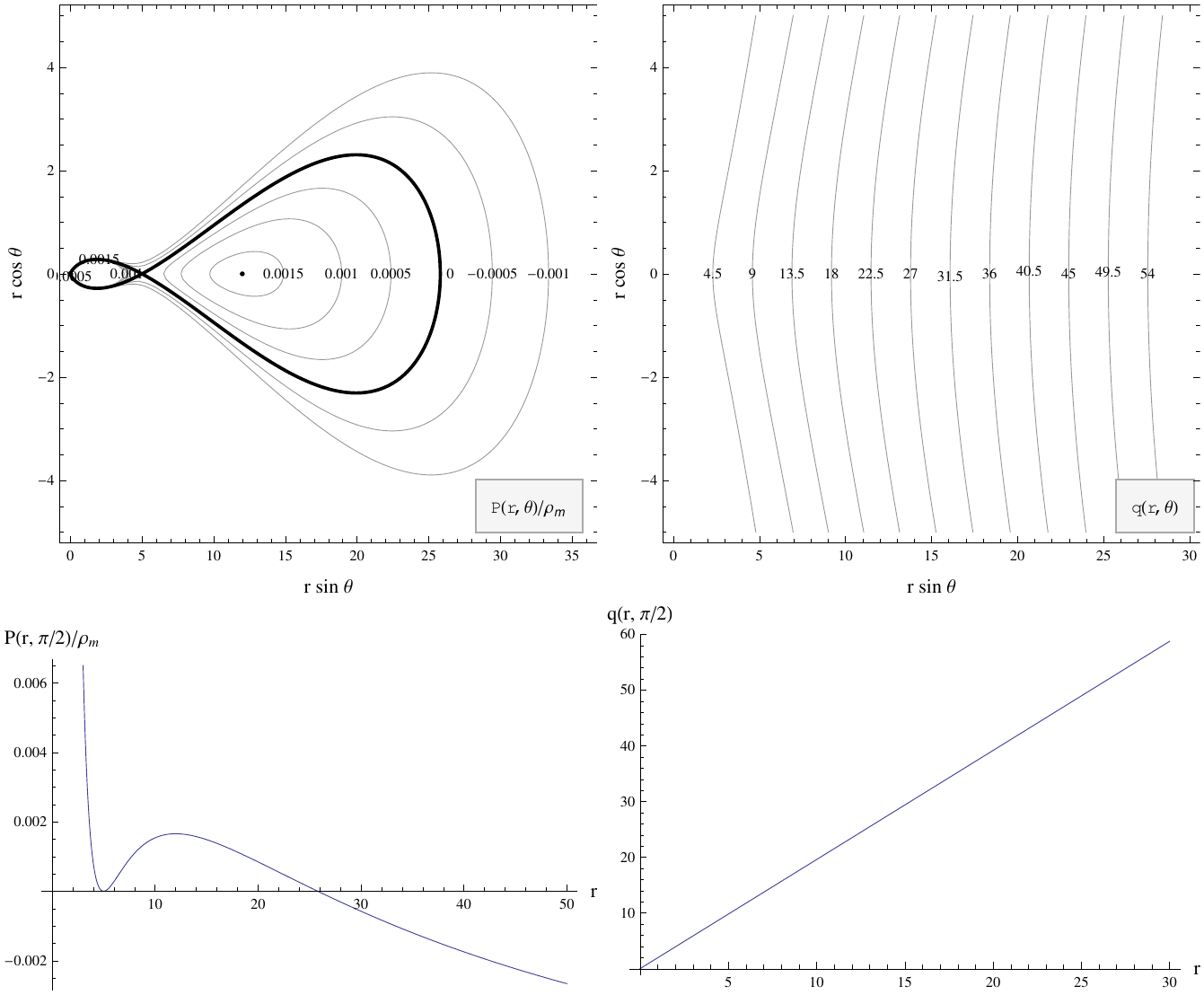}
	\caption{Positively charged equatorial torus (\emph{family III}). The meridional section and equatorial profile of the pressure field (left) and the specific electric charge (right) are shown.}
	\label{f6}
\end{figure*}

\subsubsection{Family I equatorial tori}
As an example, we choose $K_1=-1$, which corresponds to the torus with uniform distribution of the specific angular momentum, $\ell(r,\,\theta)=K_2=\mbox{const}$. In this case, the specific charge distribution and the pressure field are given by the relations
\be                                                                 \label{e34}
q(r,\,\theta)=Cr^3,
\ee
\be                                                                 \label{e35}
P(r,\,\theta)=P_0+\rho_{\rm m}\frac{GM}{r}-\frac{1}{2}\rho_{\rm m}v_{\phi}^2+\rho_{\rm m}K_2 \mu C\ln\left(\frac{\sin^2\theta}{r}\right),
\ee
where
\be                                                                 \label{e36}
v_{\phi}=\frac{K_2}{r\sin\theta},
\ee
and $P_0=\mbox{const}$ is an integration constant, determining the surface of zero pressure (boundary of the torus) and, thus, the torus extension and thickness as well. 

To obtain a particular pressure distribution, we further choose $P_0=0$, $K_{2}/\sqrt{GM}=2.5$, and, in accordance with conditions (\ref{e25}) and (\ref{e26}) graphically presented for the case $K_1=-1$ in Fig.~\ref{f1}, $r_{\rm c}=6$. Further from relation (\ref{e27}) we obtain $\mu C/\sqrt{GM}\doteq 2.78\times 10^{-3}$, which corresponds to the positively charged torus ($C>0$). The meridional section and equatorial profile of the resulting pressure field, as well as the specific charge distribution, are presented in Fig.~\ref{f2}.\footnote{In all figures, units $G=M=1$ are used.} We see that there are toroidal isobaric surfaces corresponding to equilibrium stationary fluid torus with the spherical distribution of a positive specific charge. 

The negatively charged torus is determined by the condition $C<0$ ($C$ is given by relation (\ref{e27})) which, together with conditions (\ref{e25}) and (\ref{e26}), puts a~limit on the location of the torus center. For the given values of $K_1=-1$ and $K_2/\sqrt{GM}=2.5$, the center of the negatively charged torus must be located in between the radii $6.25$ and $9.375$, see Fig.~\ref{f1}. We choose $r_{\rm c}=8$, which gives $\mu C/\sqrt{GM}\doteq -0.012$. In this case, the equatorial pressure profile has two local extrema (maximum and minimum) around which the toroidal isobaric surfaces are centered. Nevertheless, only those around local maximum correspond to toroidal fluid body with pressure gradient decreasing from the center to an outer edge, which is, in this case, given by the critical isobaric surface, $P_{\rm crit}=0$, with two cusps out of the equatorial plane ($P_0\doteq -0.13$ in this case). The cusps enable outflow of matter from the torus to an outer space due to the violation of hydrostatic equilibrium (the so-called Paczy\'{n}ski mechanism known in the theory of accretion disks). The meridional section and equatorial profile of the pressure field, as well as the specific charge distribution, are presented in Fig.~\ref{f3}.

\begin{figure*}
\centering
  \includegraphics[width=1\hsize]{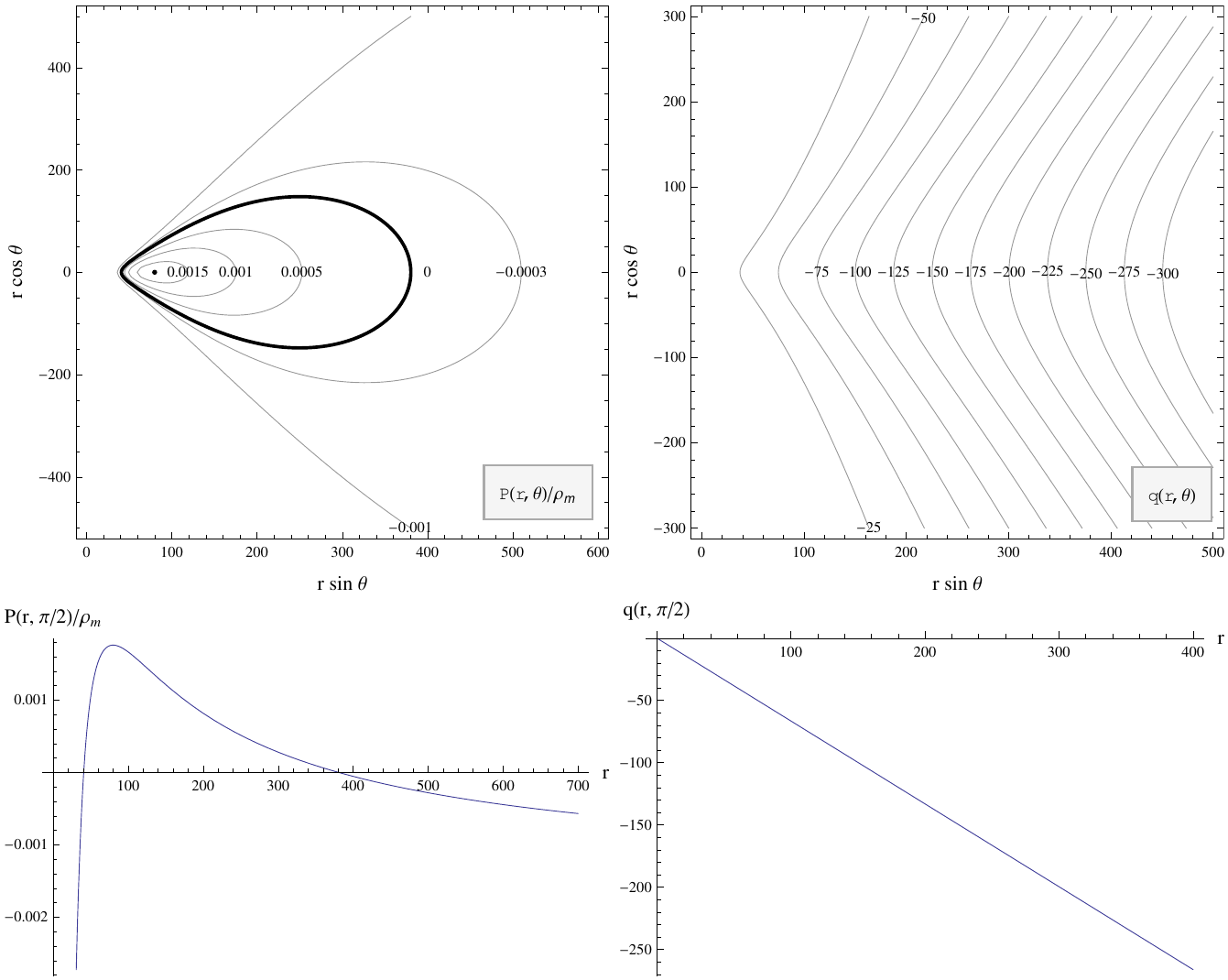}
	\caption{Negatively charged equatorial torus (\emph{family III}). The meridional section and equatorial profile of the pressure field (left) and the specific electric charge (right) are shown.}
	\label{f7}
\end{figure*}

\subsubsection{Family II equatorial tori}
Again, we choose $K_1=-1$. The corresponding specific charge distribution and pressure field are described by the formulae
\be                                                                 \label{e37}
q(r,\,\theta)=Cr^{3/2}\sin^3\theta,
\ee
\be                                                                 \label{e38}
P(r,\,\theta)=P_0+\rho_{\rm m}\frac{GM}{r}-\frac{1}{2}\rho_{\rm m}v_{\phi}^2 +\frac{2}{3}\rho_{\rm m}K_2 \mu C\left(\frac{\sin^2\theta}{r}\right)^{3/2},
\ee
where
\be                                                                 \label{e39}
v_{\phi}=\frac{K_2}{r\sin\theta},
\ee
$K_2=\ell=\mbox{const}$, and $P_0=\mbox{const}$. 

Further, we choose $P_0=-0.02$, $K_{2}/\sqrt{GM}=2.5$ and, in accordance with condition (\ref{e29}) depicted for the case $K_1=-1$ in Fig.~\ref{f1}, $r_{\rm c}=6$. According to relation (\ref{e30}), $\mu C/\sqrt{GM}\doteq 0.041$. The meridional section and equatorial profile of the pressure field, as well as the specific charge distribution, are presented in Fig.~\ref{f4}. The relevant part of the pressure field has a~toroidal topology, while the specific charge distribution has a~cylindrical topology. The presented case corresponds to a~positively charged torus. 

The negatively charged torus with the same values of $K_1$ and $K_2$ would be centered around a circle with the radius $6.25<r_{\rm c}<9.375$, see Fig.~\ref{f1}. The equatorial pressure profile of negatively charged tori still has only one local extreme (maximum) and the pressure field has the same topology as in the positively charged case.

\begin{figure}
\centering
  \includegraphics[width=1 \hsize]{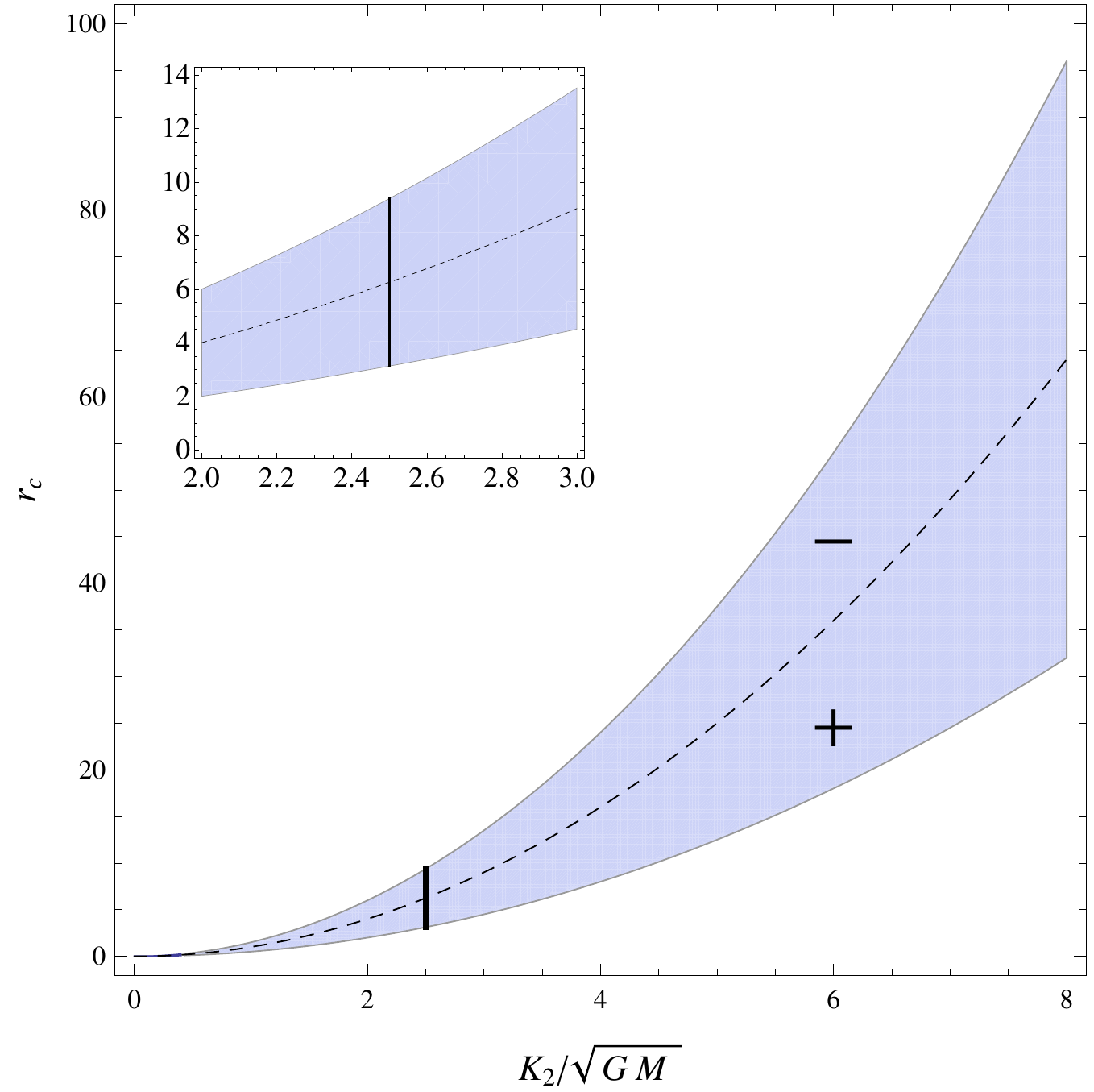}
  \caption{Region of parameters $K_2$ and $r_{\rm c}$, enabling the existence of family IV equatorial tori with a~uniform distribution of the specific angular momentum, $\ell(r,\,\theta)=K_2=\mbox{const}$. The dashed curve divides the region into subregions with positively/negatively ($+/-$) charged tori. The interval of possible values of $r_{\rm c}$ for the choice $K_{2}/\sqrt{GM}=2.5$ is marked and enlarged.}
  \label{f8}
\end{figure}

\subsubsection{Family III equatorial tori}
Here, we choose $K_1=-2/3$.\footnote{The choice $K_1=-1$ used in previous examples gives the specific charge distribution of the last example and thus leads to the pressure field described by formula (\ref{e38}).} The corresponding specific charge distribution and pressure field are described by the formulae
\be                                                                 \label{e40}
q(r,\,\theta)=Cr\sin^3\theta,
\ee
\be                                                                 \label{e41}
P(r,\,\theta)=P_0+\rho_{\rm m}\frac{GM}{r}-\frac{3}{4}\rho_{\rm m}v_{\phi}^2 +\frac{3}{5}\rho_{\rm m}K_2 \mu C\left(\frac{\sin^2\theta}{r}\right)^{5/3},
\ee
where $P_0=\mbox{const}$,
\be                                                                 \label{e42}
v_{\phi}=\frac{K_2}{(r\sin\theta)^{2/3}},
\ee
and $K_2=\mbox{const}$. The distribution of the specific angular momentum is given by the relation
\be                                                                 \label{e43}
\ell(r,\,\theta)=K_2(r\sin\theta)^{1/3}.
\ee

Further, we choose $K_2/\sqrt{GM}=2.0$ and, in accordance with conditions (\ref{e31}) and (\ref{e32}) graphically presented for the case $K_1=-2/3$ in Fig.~\ref{f5}, $r_{\rm c}=12$. These choices give positively charged torus, since according to relation (\ref{e33}) $\mu C/\sqrt{GM}\doteq 1.958$. The structure of isobars, the equatorial radial profile of the pressure, and the specific charge distribution in the fluid are shown in Fig.~\ref{f6}. We see that there is one self-crossing isobar of the `Roche-lobe' type surrounding closed isobars and forming a cusp that enables the outflow of fluid from the torus onto the central body. Note that the constant $P_0$ is chosen to obtain $P_{\rm crit}=0$ for this critical isobar. The location of the cusp corresponds to the position of a~local minimum in the radial pressure profile. The surfaces of constant specific charge have cylindrical topology.

In order to obtain a negatively charged torus with the same values of $K_1$ and $K_2$ as in the case of positively charged torus presented above, we must shift a~torus center to the position $64<r_{\rm c}<216$, see Fig.~\ref{f5}. We choose $r_{\rm c}=80$, which gives $\mu C/\sqrt{GM}\doteq -0.665$, and $P_0=-0.0015$. The resulting pressure and specific charge distributions are shown in Fig.~\ref{f7}. There is no self-crossing isobar, only the toroidal ones exist, corresponding to a~negatively charged toroidal structure with the cylindrical distribution of the electric charge.

\begin{figure*}
\centering
  \includegraphics[width=1 \hsize]{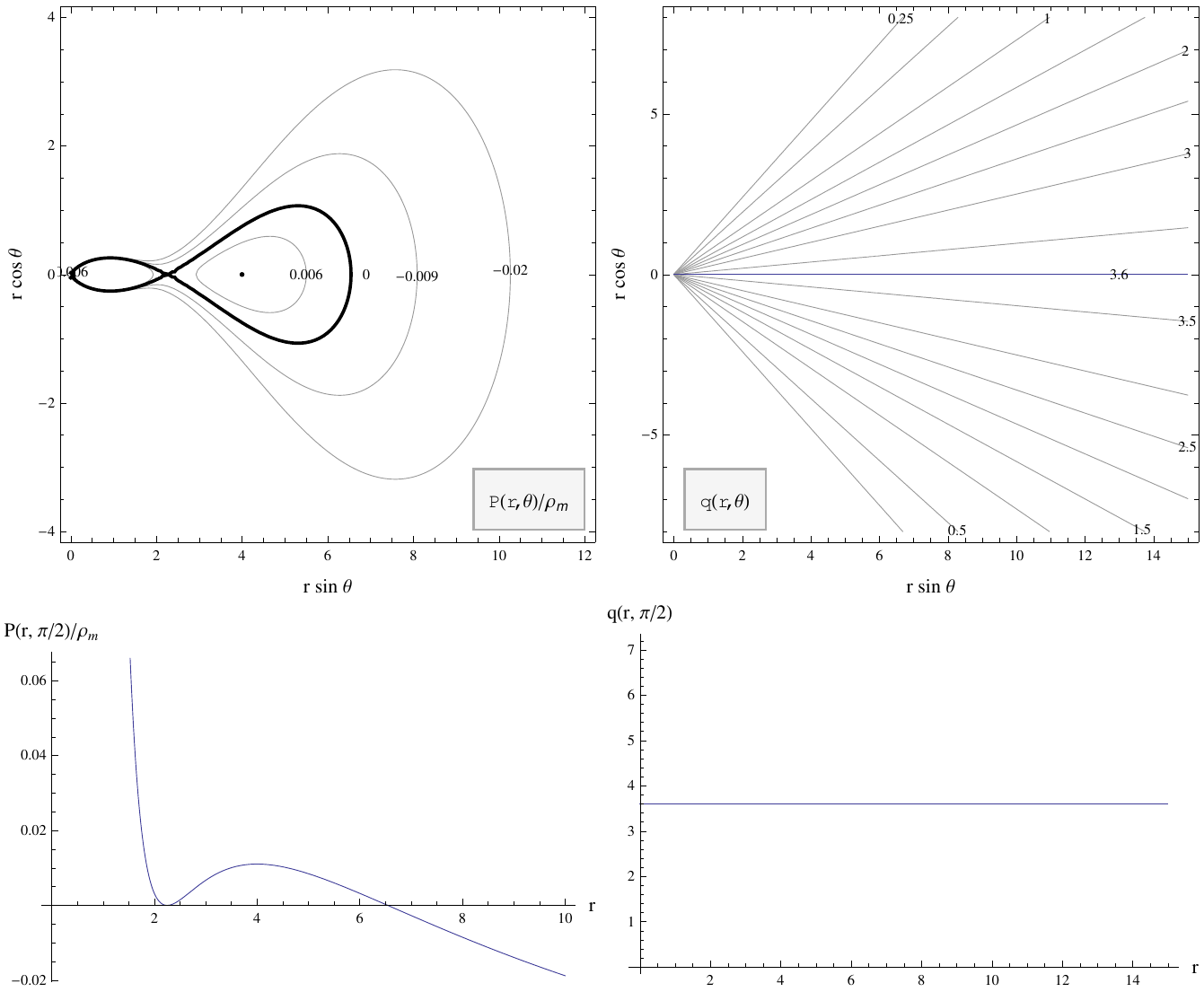}
	\caption{Positively charged equatorial torus (\emph{family IV}). The meridional section and equatorial profile of the pressure field (left) and the specific electric charge (right) are shown.}
	\label{f9}
\end{figure*}

\subsubsection{Family IV equatorial tori}
As an example, again we choose tori with uniform distribution of the specific angular momentum, $\ell(r,\,\theta)=\mbox{const}$, which are characterized by $K_1=-1$. The corresponding specific charge distribution and the pressure field are described by the formulae
\be                                                                 \label{e43.1}
q(r,\,\theta)=C\sin^6\theta,
\ee
\be                                                                 \label{e43.2}
P(r,\,\theta)=P_0+\rho_{\rm m}\frac{GM}{r}-\frac{1}{2}\rho_{\rm m}v_{\phi}^2 +\frac{1}{3}\rho_{\rm m}K_2 \mu C\left(\frac{\sin^2\theta}{r}\right)^{3},
\ee
where
\be                                                                 \label{e43.3}
v_{\phi}=\frac{K_2}{r\sin\theta},
\ee
$K_2=\ell=\mbox{const}$, and $P_0=\mbox{const}$. 

For a particular pressure distribution, we choose $K_{2}/\sqrt{GM}=2.5$ and $r_{\rm c}=4$ in accordance with conditions (\ref{e33.1}) and (\ref{e33.2}) represented for the case $K_1=-1$ in Fig.~\ref{f8}. These choices give a~positively charged torus, since, according to relation (\ref{e33.3}), $\mu C/\sqrt{GM}=3.6$. The structure of isobars, the equatorial radial profile of the pressure, and the specific charge distribution in the fluid are shown in Fig.~\ref{f9}. As in the case of family III tori, there is one self-crossing isobar of the `Roche-lobe' type surrounding closed isobars and forming a cusp (the constant $P_0$ is chosen to obtain $P_{\rm crit}=0$ for this critical isobar). The location of the cusp is given by the position of the local minimum in the radial pressure profile. The surfaces of constant specific charge correspond to vertical planes.

\begin{figure*}
\centering
  \includegraphics[width=1\hsize]{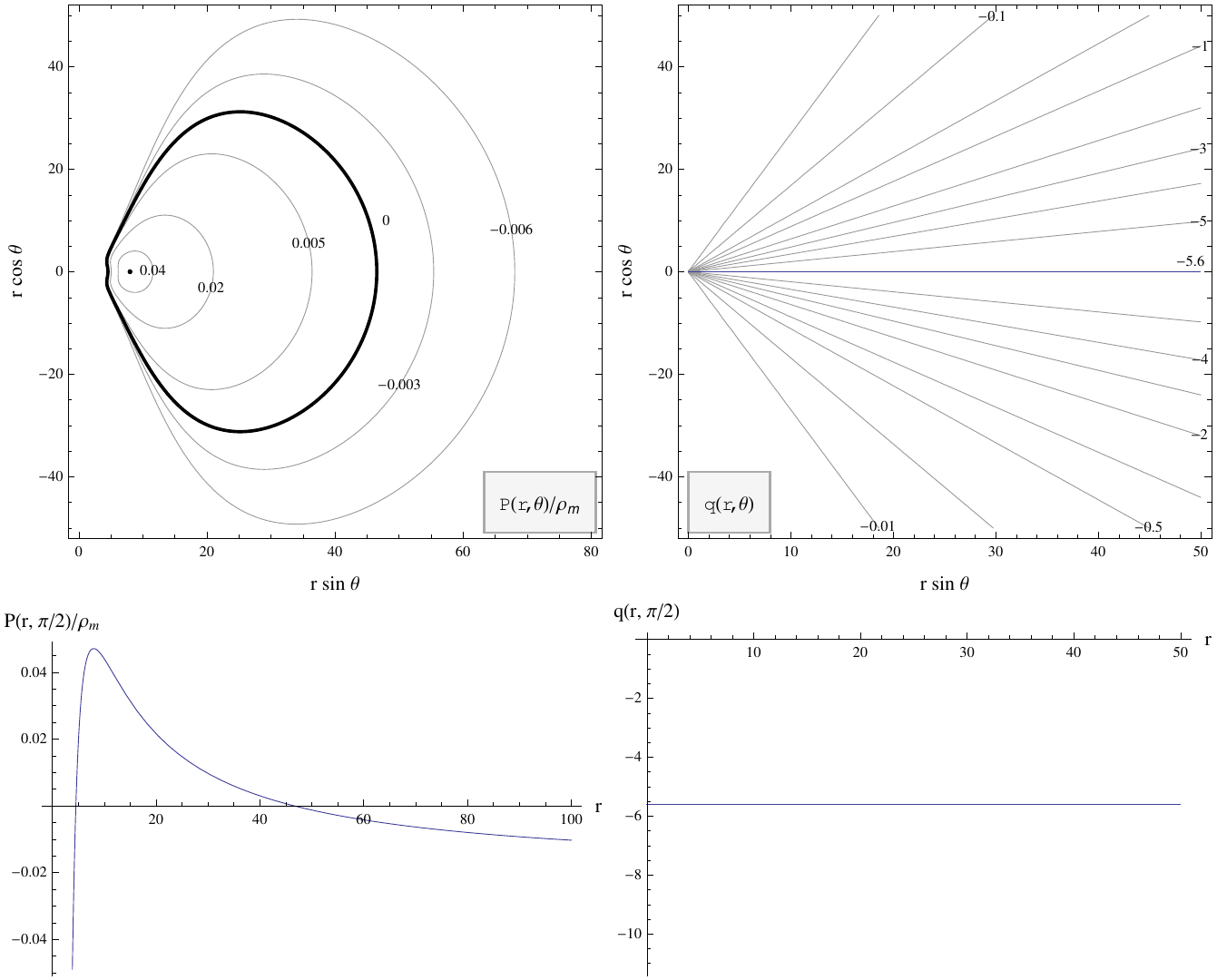}
	\caption{Negatively charged equatorial torus (\emph{family IV}). The meridional section and equatorial profile of the pressure field (left) and the specific electric charge (right) are shown.}
	\label{f10}
\end{figure*}

The negatively charged torus with the same values of $K_1$ and $K_2$ as the positively charged one presented above must be centered around a circle with the radius $6.25<r_{\rm c}<9.375$, see Fig.~\ref{f8}. We choose $r_{\rm c}=8$, which gives $\mu C/\sqrt{GM}=-5.6$, and $P_0=-0.02$. Resulting pressure and specific charge distributions are shown in Fig.~\ref{f10}. There is no self-crossing isobar, only the toroidal ones exist, corresponding to a~negatively charged toroidal structure with the radial distribution of the electric charge.

\subsection{Off-equatorial tori}
Due to the mutual interaction between electric charges and the magnetic field, the torus center can be lifted away from the equatorial plane, i.e., $\theta_{\rm c}\neq\pi/2$. The precise location of the center is governed by the interplay with pressure gradients. Since the specific charge is given by relation (\ref{e21}) there, we immediately see that for positively/negatively rotating torus ($v_{\phi}>0/v_{\phi}<0$) and the positive/negative direction of external magnetic field ($\mu>0/\mu<0$), the off-equatorial torus must possess a~negative electric charge. On the other hand, positively/negatively rotating tori in negatively/positively oriented magnetic fields possess a~positive electric charge. Next we will focus only on off-equatorial tori with $v_{\phi}>0$ orbiting in the external magnetic field characterized by $\mu>0$, i.e., on the negatively charged off-equatorial tori.

In the torus center, there is $\nabla P=0$. Adding relation (\ref{e21}) for the specific charge to Eq.~(\ref{e12}) and putting $\p P/\p r =0$, we obtain the relation for the orbital velocity in the center
\be                                                                 \label{e44}
v_\phi(r_{\rm c},\,\theta_{\rm c})=\sqrt{\frac{2GM}{3r_{\rm c}}}.
\ee 
Consequently, relation (\ref{e21}) for the specific charge in the center takes the form
\be                                                                 \label{e45}
q(r_{\rm c},\,\theta_{\rm c})=-\frac{1}{2\mu}\sqrt{\frac{2GM}{3r_{\rm c}}}\frac{r_{\rm c}^2}{\sin\theta_{\rm c}}.
\ee 

Analysis of conditions (\ref{e23}) shows that off-equatorial local maxima of the pressure can exist for families II, III and IV of the specific charge distribution $q(r,\,\theta)$. In detail, in the torus center the constants $K_1$, $K_2$, and $C$ must satisfy conditions
\begin{description}
\item[family II]
	\bea
	K_1 &<& -\frac{1}{2},                                             \label{e46} \\
	\mu C &=& -\frac{K_2}{2}r_{\rm c}^{(2K_1+1)/2}(\sin\theta_{\rm c})^{4K_1-1}. \label{e47}
	\eea
\item[family III]
  \bea
  K_1 &\in& (-3.26,\,-0.41),                                        \label{e48} \\
  \mu C &=& -\frac{K_2}{2}r_{\rm c}^{(5K_1+4)/2}(\sin\theta_{\rm c})^{K_1-4}. \label{e49}
  \eea	
\item[family IV]
  \bea
  K_1 &<& -\frac{\sqrt{10}-2}{3}\doteq -0.387,                       \label{e49.1} \\
  \mu C &=& -\frac{K_2}{2}r_{\rm c}^{(K_1+2)}(\sin\theta_{\rm c})^{4(K_1-1)}. \label{e49.2}
  \eea	
\end{description}

Further, for each of these three families, we present an~illustrative example of off-equatorial toroidal structures. Again, we must specify the integration constants $K_1$, $K_2$, and $C$. However, instead of $C$, it is more convenient to choose the latitude of the torus center, i.e., $\theta_{\rm c}$.

\begin{figure*}
\centering
  \includegraphics[width=1\hsize]{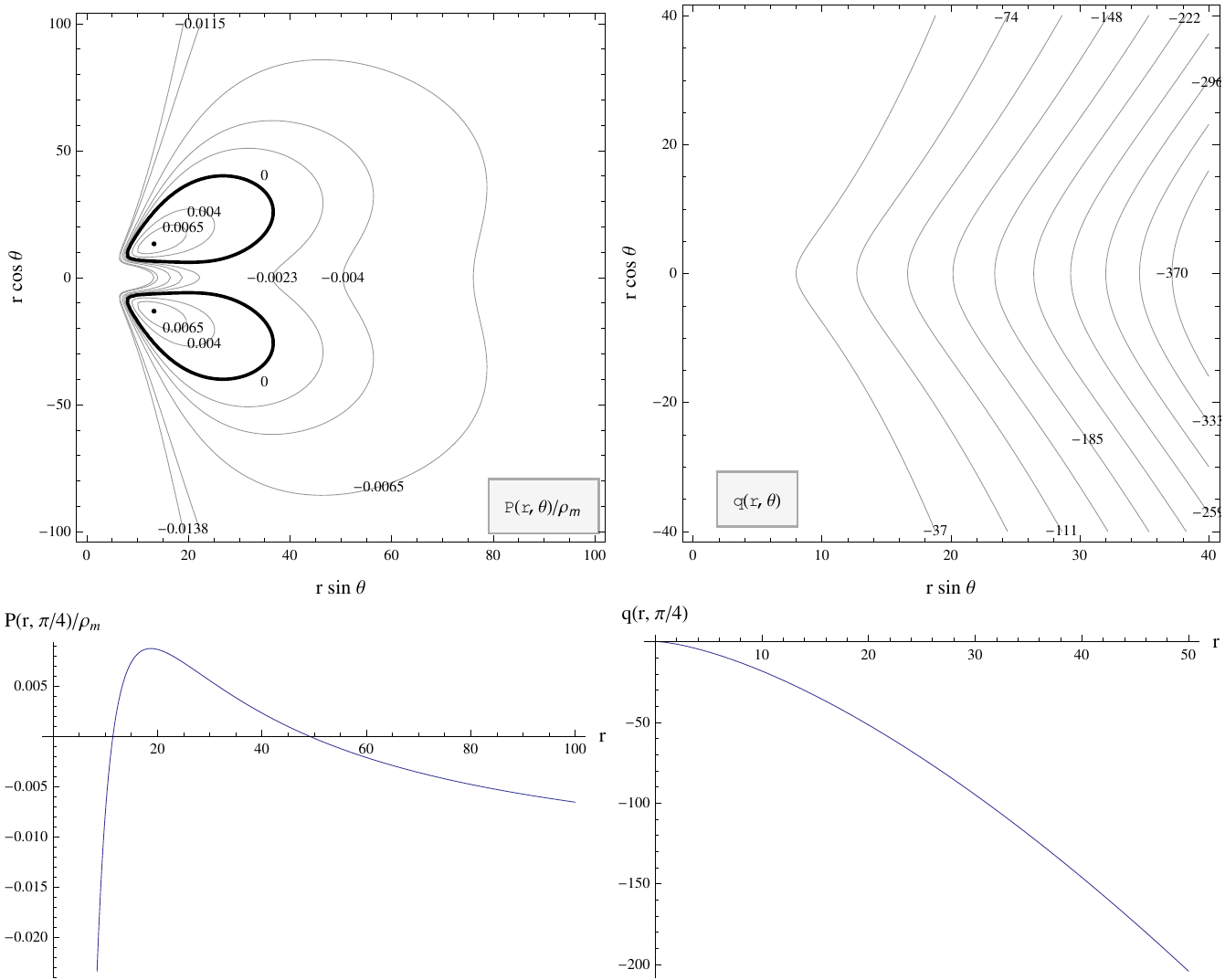}
	\caption{Negatively charged off-equatorial tori (\emph{family II}). The meridional section and off-equatorial ($\theta=\pi/4$) profile of the pressure field (left) and the specific electric charge (right) are shown.}
	\label{f11}
\end{figure*}

\subsubsection{Family II off-equatorial tori}

In order to be consistent with previously studied equatorial tori, we choose the family II tori with uniform distribution of the specific angular momentum, $\ell(r,\,\theta)=K_2=\mbox{const}$. In this case, $K_1=-1$ and the specific charge, pressure and orbital velocity fields are given by relations (\ref{e37})--(\ref{e39}). Further, we choose $P_0=-0.015$, $K_{2}/\sqrt{GM}=2.5$, and $\theta_{\rm c}=\pm\pi/4$ (as the pressure field is symmetric with respect to the equatorial plane). The radial coordinate of tori centers is determined by relations (\ref{e39}) and (\ref{e44}) for orbital velocity, which have to be fulfilled simultaneously in the center. The chosen values of $K_2$ and $\theta_{\rm c}$ give $r_{\rm c}=18.75$ and, according to (\ref{e47}), $\mu C/\sqrt{GM}\doteq -1.633$. The meridional sections and off-equatorial ($\theta=\pi/4$) profiles of both the pressure field and specific charge distribution are presented in Fig.~\ref{f11}, showing toroidal off-equatorial structures above and under the equatorial plane with cylindrical distribution of the specific charge.

\subsubsection{Family III off-equatorial tori}

\begin{figure*}
\centering
  \includegraphics[width=1\hsize]{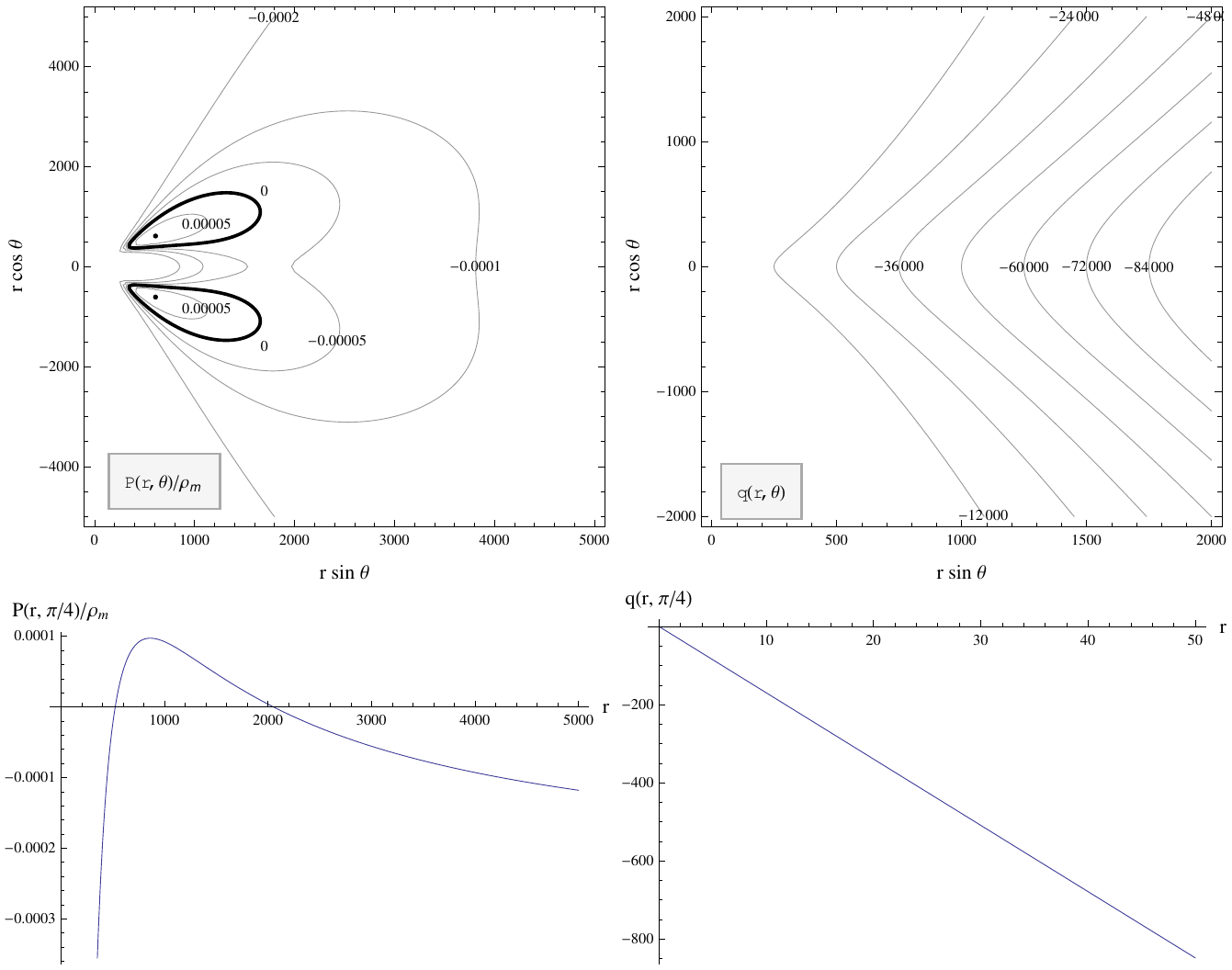}
  \caption{Negatively charged off-equatorial tori (\emph{family III}). The meridional section and off-equatorial ($\theta=\pi/4$) profile of the pressure field (left) and the specific electric charge (right) are shown.}
	\label{f12}
\end{figure*}

As in the case of equatorial family III tori, we choose $K_1=-2/3$. The specific charge distribution, pressure field, orbital velocity field and specific angular momentum distribution are thus given by relations (\ref{e40})--(\ref{e43}). Further, we choose $K_{2}/\sqrt{GM}=2.0$ and $\theta_{\rm c}=\pm\pi/4$, giving the radial coordinate of tori centers $r_{\rm c}=864$. Subsequently, $\mu C/\sqrt{GM}= -48$. For $P_0=-2.5\times 10^{-4}$, the meridional sections and off-equatorial ($\theta=\pi/4$) profiles of both the pressure field and specific charge distribution are presented in Fig.~\ref{f12}. Again, there are toroidal off-equatorial structures located symmetrically above and under the equatorial plane, having cylindrical distribution of the specific charge.

\subsubsection{Family IV off-equatorial tori}

\begin{figure*}
\centering
  \includegraphics[width=.92\hsize]{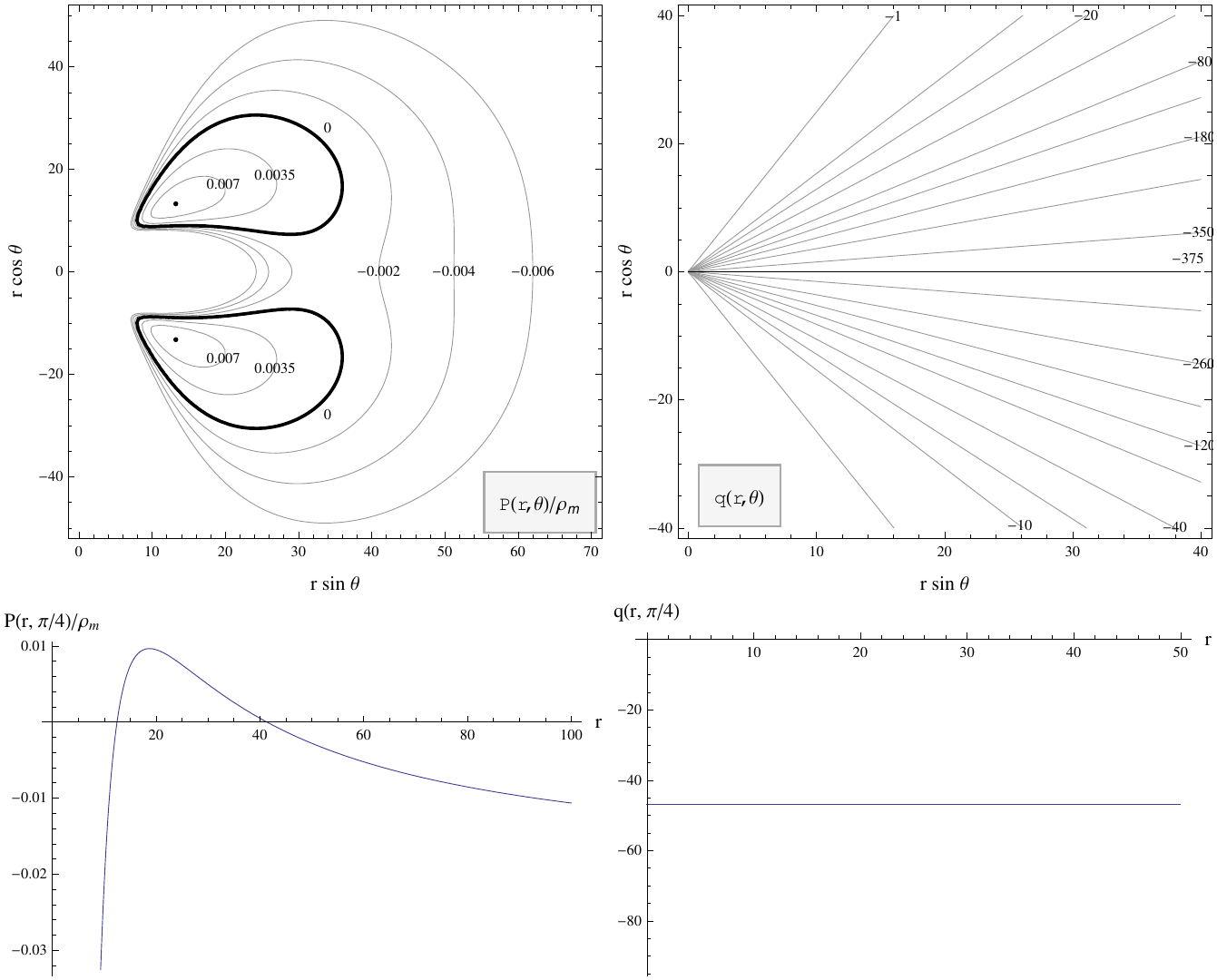}
  \caption{Negatively charged off-equatorial tori (\emph{family IV}). The meridional section and off-equatorial ($\theta=\pi/4$) profile of the pressure field (left) and the specific electric charge (right) are shown.}
	\label{f13}
\vskip3ex	
\includegraphics[width=.92\hsize]{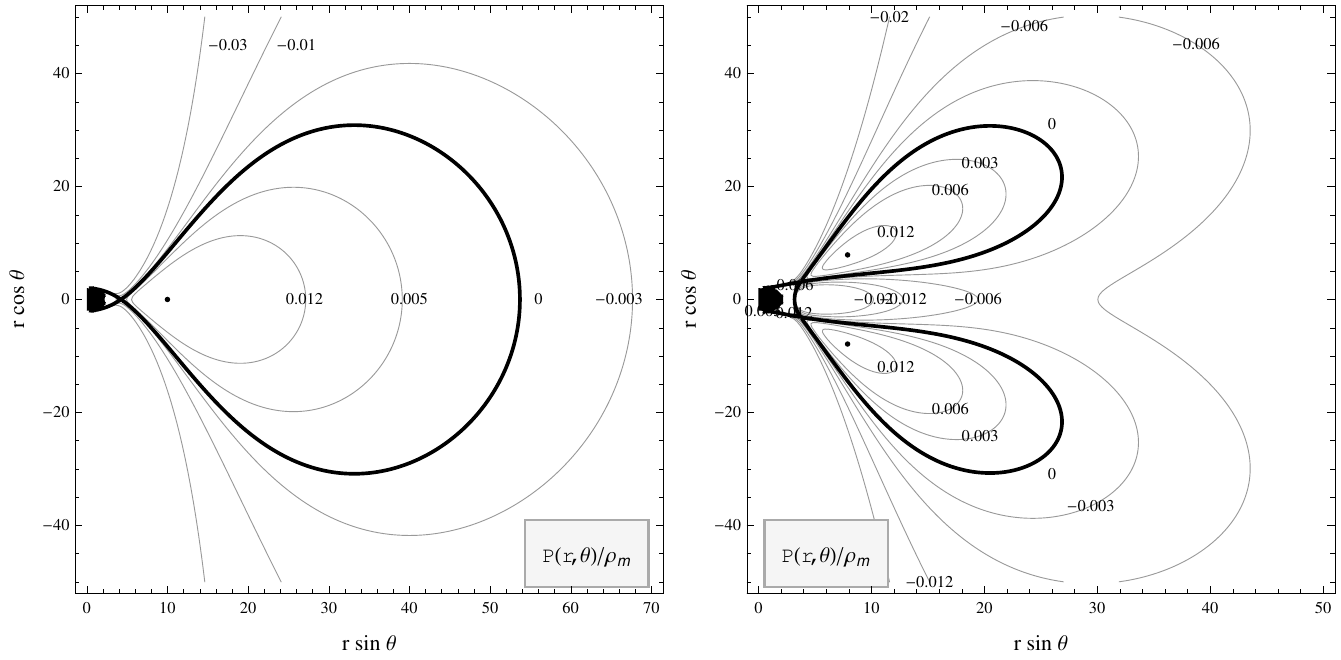}
	\caption{Negatively charged Paczy\'{n}ski--Wiita equatorial and off-equatorial tori (\emph{family II}). The boundary of each torus is given by a~marginally closed isobar forming the cusp which enables accretion onto the central object represented by a black circle of Schwarzschild radius.}
	\label{f14}
\end{figure*}

As in the case of equatorial family IV tori, we choose $K_1=-1$. The specific charge distribution, pressure field and orbital velocity field are thus given by relations (\ref{e43.1})--(\ref{e43.3}). Further, we choose $K_{2}/\sqrt{GM}=2.5$ and $\theta_{\rm c}=\pm\pi/4$, giving the radial coordinate of tori centers $r_{\rm c}=18.75$. Subsequently, $\mu C/\sqrt{GM}= -375$. For $P_0=-0.02$, the meridional sections and off-equatorial ($\theta=\pi/4$) profiles of both the pressure field and specific charge distribution are presented in Fig.~\ref{f13}. There are toroidal off-equatorial structures located symmetrically above and under the equatorial plane, now having a~radial distribution of the specific charge.

\section{Polytropic fluid} \label{s4}
For any perfect fluid, the basic set of partial differential equations, describing the axisymmetric motion of charged fluid in spherical gravitational and dipolar magnetic fields, has the form of Eqs. (\ref{e12}) and (\ref{e13}), where $\rho_{\rm m}=\rho_{\rm m}(r,\,\theta)$ in general. Further, we define the function 
\be                                                 \label{e50}
h(r,\,\theta)=\frac{P}{\rho_{\rm m}}. 
\ee

In the case of an~incompressible fluid ($\rho_{\rm m}=\mbox{const}$), the given set of equations can be written as
\bea                                                \label{e51}
\frac{\p h}{\p r} &=& \mathcal{A}(r,\,\theta), \\   \label{e52}
\frac{\p h}{\p\theta} &=& \mathcal{B}(r,\,\theta).
\eea

Now, we take a polytropic fluid with the equation of state
\be                                                 \label{e53}
P=K\rho_{\rm m}^{\gamma}.
\ee
It is easy to show that, in this case, Eqs. (\ref{e12}) and (\ref{e13}) take the form
\bea                                                \label{e54}
\frac{\p h}{\p r} &=& \frac{\gamma-1}{\gamma}\mathcal{A}(r,\,\theta), \\   \label{e55}
\frac{\p h}{\p\theta} &=& \frac{\gamma-1}{\gamma}\mathcal{B}(r,\,\theta).
\eea
Clearly, in the case of a~polytropic fluid, solutions for the function $h$ differ from those for an incompressible fluid just by the multiplicative factor $(\gamma -1)/\gamma$, where $\gamma$ is the polytropic index. Since in the polytropic fluid surfaces of $P=\mbox{const}$ coincide with the surfaces of $\rho_{\rm m}=\mbox{const}$, it is sufficient to analyze the function $h$ in searching for stationary toroidal structures. Therefore, the discussion of incompressible perfect fluid tori presented in the previous section is fully relevant also for polytropic perfect fluid and reveals the existence of both the equatorial and off-equatorial toroidal structures also in polytropic case.

\section{Pseudo-Newtonian model} \label{s5}
The formalism given above can be used also for modeling charged toroidal structures in the vicinity of compact objects like neutron stars or black holes within the pseudo-Newtonian approach. In this case, we can use the well-known Paczy\'{n}ski--Wiita (PW) potential instead of the Newtonian one to describe the gravitational field near compact spherical body. 

The PW potential is given by the relation \citep{Pac-Wii:1980:ASTRA:}
\be                                                 \label{e56}
\Phi_{\rm PW} = -\frac{GM}{r-2R_{\rm g}},
\ee
where $R_{\rm g}\equiv GM/c^2$ is the gravitational radius of the compact object. The PW potential is able to capture some important properties of the gravitational field of compact objects following from General Relativity (e.g., locations of marginally stable and marginally bound circular orbits), which are crucial for the orbital motion of matter in their vicinity. For its `derivation' and following discussion, see \citet{Abr:2009:ASTRA:}.

In terms of the PW potential, the fluid dynamics is described by a~slightly modified condition of hydrostatic equilibrium. In the radial direction, it takes the form
\be                                                                      \label{e57}
\frac{\p P}{\p r} = -\rho_{\rm m}\frac{GM}{(r-2R_{\rm g})^2}+\rho_{\rm m}\frac{v_{\phi}^2}{r}-\rho_{\rm e}v_{\phi}\mu\frac{\sin\theta}{r^3},
\ee
while in the latitudinal direction, the condition retains the form (\ref{e10}). 

Again, we can assume relation between the charge density $\rho_{\rm e}$ and mass density $\rho_{\rm m}$ in the form (\ref{e11}). Since the results for a polytropic fluid are given as the scaled version of those for an incompressible fluid, we restrict our attention to the incompressible fluid only. 

The integrability condition satisfies the same form (\ref{e15}), leading again to the orbital velocity (\ref{e17}) and to four possible families of specific charge distribution (\ref{e18})--(\ref{e20.1}). Analysis of necessary and sufficient conditions for the existence of pressure maxima, $\nabla P=0$ and conditions (\ref{e23}), reveals that in the case of PW potential, the maxima can also be located in the equatorial plane ($\theta_{\rm c}=\pi/2$) or out of the equatorial plane ($\theta_{\rm c}\neq\pi/2$). However, apart from the local maximum, there is also a local minimum of the pressure, corresponding to the cusp in the pressure distribution, i.e., to the point in which one of the isobars crosses itself. In dependence on the rotational law, $v_{\phi}=v_{\phi}(\ell)$, this isobar can be marginally closed, forming the cusp, or it can be open. In the former case, little overflow of such isobar enables an outflow of matter from the torus through the cusp.

Next, we present illustrative examples of both the equatorial and off-equatorial toroidal structures constructed with the use of PW potential, showing the existence of cusps, but not presenting a~detailed analysis of all the possible cases.

In order to have some connection with previously presented Newtonian tori, we choose family II tori with $K_1=-1$, i.e., with a~uniform distribution of the specific angular momentum, $\ell(r,\,\theta)=\mbox{const}$. The corresponding specific charge distribution is given by relation (\ref{e37}) and the pressure field is described by the formula
\be                                                                   
P(r,\,\theta)=P_0+\rho_{\rm m}\frac{GM}{r-2R_{\rm g}}-\frac{1}{2}\rho_{\rm m}v_{\phi}^2 +\frac{2}{3}\rho_{\rm m}K_2 \mu C\left(\frac{\sin^2\theta}{r}\right)^{3/2}, \label{e58}
\ee
where the orbital velocity $v_{\phi}(r,\,\theta)$ is given by formula (\ref{e39}). In dependence on the constants $K_2$ and $C$, the torus is located in or out of the equatorial plane. Instead of $C$, the location of the torus center can be prescribed. Note that in the chosen case of uniformly distributed specific angular momentum, $\ell(r,\,\theta)=K_2$. In the case of equatorial torus, we choose $K_{2}/\sqrt{GM}=3.8$ and the position of torus center $r_{\rm c}=10$, which imply $\mu C/\sqrt{GM}\doteq -0.0986$. In the case of off-equatorial torus, we choose $K_{2}/\sqrt{GM}=2.35$ and the latitude of the torus center $\theta_{\rm c}=\pi/4$, giving $r_{\rm c}\doteq 11.16$ and $\mu C/\sqrt{GM}\doteq -1.989$. Constant $P_0$ in both the cases is chosen to obtain for the critical self-crossing isobar the zero level. The meridional section through the pressure field of both the equatorial and off-equatorial toroidal structures is shown in Fig.~\ref{f14}. 

\section{Conclusions} \label{s6}
In this paper, we presented a~Newtonian description of charged perfect fluid, i.e., a~perfect fluid whose particles carry electric charges, orbiting in spherical gravitational and dipolar magnetic fields. 
The equation of motion was analyzed with the aim of finding stationary toroidal structures---electrically charged thick disks, existing in prescribed external gravitational and magnetic fields. 

From an~astrophysical point of view, our investigation can provide insight into the interplay of gravitational and electromagnetic forces determining the vertical structure of dusty tori composed of electrically charged grains (dust) in which pressure gradients coming from mutual interactions are not negligible. We adopted a toy-model approach which is obviously too simple to describe real cosmic structures, such as those around supermassive black holes in the nuclei of galaxies. However, we suggest that our model demonstrates an interesting possibility that a small electric charge developed on the grains affects significantly their motion through gravitational and pervasive (large-scale) magnetic fields. For the former, we assumed the approximation of central Newtonian gravitating body. For the latter, we adopted an organized (dipolar) magnetic field with its axis oriented perpendicularly to the orbital plane.

It was shown that electric charge distributed in the fluid changes dramatically the properties of possible toroidal structures. Namely, apart from equatorial toroidal structures, the off-equatorial tori can also exist due to electromagnetic interaction of the disk with an external dipolar magnetic field. For the given orientation of the magnetic field ($\mu>0$) and the positive rotation of the fluid ($v_{\phi}>0$), the off-equatorial tori can be only negatively charged, while the equatorial tori can be both positively or negatively charged. Therefore, an appropriate adding of both the solutions (even for the same specific angular momentum distribution) can give an electrically neutral configuration composed of the equatorial positively charged torus and two negatively charged tori above and under the equatorial plane. 

Another interesting and qualitatively new result is the presence of marginally closed isobars with cusp(s) in terms of Newtonian gravity. Note that in terms of Einstein's gravity (General Relativity), the existence of cusps in the structure of isobars is a natural phenomenon related to specific properties of motion in curved spacetimes, leading to the existence of accreting thick disks \citep{Abr-Jar-Sik:1978:ASTRA:}. In the non-relativistic (Newtonian) case, the cusps are related to the electromagnetic interaction between the disk with charges and external magnetic field.\footnote{In the pseudo-Newtonian model with the Paczy\'{n}ski--Wiita potential, all the disks contain, at least, a cusp at the inner edge, which has relativistic, non-electromagnetic origin.} In equatorial structures, the cusps enable accretion onto the central object (disks belonging to the families III and IV) or outflows from the torus above and under the disk (the family I disks). 
However, all outflows through the cusps are driven by the same mechanism following from a violation of hydrostatic equilibrium due to small overflow of the critical isobar forming the disk surface.

A possible future extension of our work could be in the analysis of charged tori immersed into large-scale (galactic) magnetic fields. Such analysis could be further extended in order to include a~cosmic repulsion. It has been shown that the cosmic repulsion, represented by positive cosmological constant, changes substantially the structure of isobars near the so-called static radius \citep{Stu-Hle:1999:PHYSR4:}, where the cosmic repulsion is balanced by the gravitational attraction of central body.\footnote{For test-disk configurations around supermassive black holes $(10^6-10^9)\,M_{\odot}$ and current value of the cosmological constant $\Lambda_0=1.3\times 10^{-56}\,{\rm cm}^{-2}$ (in geometrical $c=G=1$ units), the static radius reaches values from tens to hundreds of kiloparsecs \citep{Stu-Sla-Hle:2000:ASTRA:,Sla-Stu:2005:CLAQG:}.} This leads to the existence of cusps at the outer edge of toroidal configurations and enables outflows of matter from such structures \citep{Stu-Sla-Hle:2000:ASTRA:,Sla-Stu:2005:CLAQG:,Kuc-Sla-Stu:2011:JCAP:}. 
The combined influence of large-scale magnetic fields and cosmic repulsion can be well studied in the pseudo-Newtonian framework by the use of cosmological Paczy\'{n}ski--Wiita potential \citep{Stu-Kov:2008:INTJMD:}, reflecting with high precision properties of fluid tori \citep{Stu-Sla-Kov:2009:CLAQG:} and interacting objects on galactic scales \citep{Stu-Sch:2011:JCAP:,Stu-Sch:2012:INTJMD:}.

\acknowledgments

P.S., J.K., and Z.S. express their gratitude for the Institutional support
of the Faculty of Philosophy and Science of Silesian University in Opava and
for the support from the project Synergy CZ.1.07/2.3.00/20.0071. J.K. thanks the Czech Science Foundation GA \v{C}R for the project P209/10/P190 and V.K. acknowledges the Czech-German (DFG-GACR) international project 13-00070J.



\end{document}